\documentclass[reprint,
 amsmath,amssymb,
 aps,
 prl,
 floatfix
]{revtex4-1}

\usepackage[
  colorlinks=true,
  urlcolor=blue,
  linkcolor=blue,
  citecolor=blue
]{hyperref}

\usepackage{graphicx}
\usepackage{dcolumn}
\usepackage{bm}
\usepackage{hyperref}
\usepackage{braket}
\usepackage{longtable}
\usepackage[usenames, dvipsnames]{color}
\usepackage{afterpage}

\definecolor{mygray}{gray}{0.6}

\newcommand{\Exp}[1]{e^{#1}}
\newcommand{\wb}{\overline{\omega}_T}
\newcommand{\wt}{\widetilde{\omega}_T}
\newcommand{\wpump}{\omega_p}

\newcommand{\phib}{\overline{\Phi}}
\newcommand{\phit}{\widetilde{\Phi}}
\newcommand{\BJ}{\mathrm{J}}

\newcommand{\units}[1]{\,\mathrm{#1}}

\begin{document}

\title{Parametrically Activated Entangling Gates Using Transmon Qubits}

\author{S.~A.~Caldwell}
\thanks{These authors contributed equally.}
\author{N.~Didier}
\thanks{These authors contributed equally.}
\author{C.~A.~Ryan}
\thanks{These authors contributed equally.}
\author{E.~A.~Sete}
\thanks{These authors contributed equally.}

\author{A.~Hudson}
\author{P.~Karalekas}
\author{R.~Manenti}
\author{M.~P.~da~Silva}
\author{R.~Sinclair}

\author{E.~Acala}
\author{N.~Alidoust}
\author{J.~Angeles}
\author{A.~Bestwick}
\author{M.~Block}
\author{B.~Bloom}
\author{A.~Bradley}
\author{C.~Bui}
\author{L.~Capelluto}
\author{R.~Chilcott}
\author{J.~Cordova}
\author{G.~Crossman}
\author{M.~Curtis}
\author{S.~Deshpande}
\author{T.~El~Bouayadi}
\author{D.~Girshovich}
\author{S.~Hong}
\author{K.~Kuang}
\author{M.~Lenihan}
\author{T.~Manning}
\author{A.~Marchenkov}
\author{J.~Marshall}
\author{R. Maydra}
\author{Y.~Mohan}
\author{W.~O'Brien}
\author{C.~Osborn}
\author{J.~Otterbach}
\author{A.~Papageorge}
\author{J.-P.~Paquette}
\author{M.~Pelstring}
\author{A.~Polloreno}
\author{G.~Prawiroatmodjo}
\author{V.~Rawat}
\author{M.~Reagor}
\author{R.~Renzas}
\author{N.~Rubin}
\author{D.~Russell}
\author{M.~Rust}
\author{D.~Scarabelli}
\author{M.~Scheer}
\author{M.~Selvanayagam}
\author{R.~Smith}
\author{A.~Staley}
\author{M.~Suska}
\author{N.~Tezak}
\author{D.~C.~Thompson} 
\author{T.-W.~To}
\author{M.~Vahidpour}
\author{N.~Vodrahalli}
\author{T.~Whyland}
\author{K.~Yadav}
\author{W.~Zeng}

\author{C.~Rigetti}

\affiliation{%
Rigetti Computing,
775 Heinz Avenue, Berkeley, CA 94710
}%

\date{\today}

\begin{abstract}
We describe and implement a family of entangling gates activated by radio-frequency flux modulation applied to a tunable transmon that is statically coupled to a neighboring transmon. The effect of this modulation is the resonant exchange of photons directly between levels of the two-transmon system, obviating the need for mediating qubits or resonator modes and allowing for the full utilization of all qubits in a scalable architecture. The resonance condition is selective in both the frequency and amplitude of modulation and thus alleviates frequency crowding. We demonstrate the use of three such resonances to produce entangling gates that enable universal quantum computation: one $i$SWAP gate and two distinct controlled Z gates. We report interleaved randomized benchmarking results indicating gate error rates of 6\% for the $i$SWAP (duration 135ns) and 9\% for the controlled Z gates (durations $175\units{ns}$ and $270\units{ns}$), limited largely by qubit coherence.
\end{abstract}

\maketitle

A central challenge in building a scalable quantum computer with superconducting qubits is the execution of high-fidelity, two-qubit gates within an architecture containing many resonant elements. As more elements are added, or as the multiplicity of couplings between elements is increased, the frequency space of the design becomes crowded and device performance suffers. In architectures composed of transmon qubits~\cite{Koch}, there are two main approaches to implementing two-qubit gates. The first utilizes fixed-frequency qubits with static couplings where the two-qubit operations are activated by applying transverse microwave drives~\cite{Blais,Chad,Rigetti,Chow,Leek,Poletto,Sheldon}. While fixed-frequency qubits generally have long coherence times, this architecture requires satisfying stringent constraints on qubit frequencies and anharmonicities~\cite{Chow,Leek,Sheldon} which requires some tunability to scale to many qubits~\cite{Hutchings2017}. The second approach relies on frequency-tunable transmons, and two-qubit gates are activated by tuning qubits into and out of resonance with a particular transition~\cite{Strauch,Dicarlo,Yamamoto2010, Kelly,Reed2013,Kelly2015,Sete2016}. However, tunability comes at the cost of additional decoherence channels, thus significantly limiting coherence times~\cite{Kerman}. In this approach the delivery of shaped unbalanced control signals poses a challenge~\cite{Kelly2015}. Such gates are furthermore sensitive to frequency crowding---avoiding unwanted crossings with neighboring qubit energy levels during gate operations limits the flexibility and connectivity of the architecture.

An alternative to these approaches is to modulate a circuit's couplings or energy levels at a frequency corresponding to the detuning between particular energy levels of interest~\cite{Bertet,Niskanen2006,Niskanen2007,BeaudoinPRA12,StrandPRB13,Nico,Vijay,McKayPRApp16,Schuster}. This enables an entangling gate between a qubit and a single resonator~\cite{BeaudoinPRA12,StrandPRB13}, a qubit and many resonator modes~\cite{Schuster}, two transmon qubits coupled by a tunable mediating qubit~\cite{McKayPRApp16,Sete2016}, or two tunable transmons coupled to a mediating resonator~\cite{Nico, Vijay}.

Building on these earlier results, we implement two entangling gates, $i$SWAP and controlled Z (CZ), between a flux-tunable transmon and a fixed-frequency transmon. The gates are activated by modulating the tunable qubit in resonance with particular pairs of transmon states: $\ket{10}$ and $\ket{01}$ for $i$SWAP and $\ket{11}$ and $\ket{20}$ (or $\ket{02}$) for CZ. This direct qubit-qubit interaction provides each maximally entangling operation in a single step, without sacrificing any qubits to a mediating role. The interaction is a first-order process with an effective strength of roughly half the qubit-qubit coupling, and the resonance conditions depend on both the frequency and amplitude of modulation. The availability of a second dimension in the control space, combined with the selectivity of the resonances, mitigates the effect of frequency crowding in architectures where each qubit has many neighbors. The modulation amplitude can be used to vary the frequency spectrum of the resonances, and up to three entangling gates, per neighbor, can be realized at each amplitude. The ability to apply both $i$SWAP and CZ between pairs of qubits may reduce circuit complexity, as these operations are not locally equivalent. However, the $i$SWAP is locally equivalent to a CNOT followed by the interchange of the qubit labels~\cite{SchuchPRA03}. Both the $i$SWAP and the CZ gates are universal for quantum computation when combined with local operations on the individual qubits~\cite{SchuchPRA03,BarencoPRA95}.

\begin{figure}
\includegraphics[width=\columnwidth]{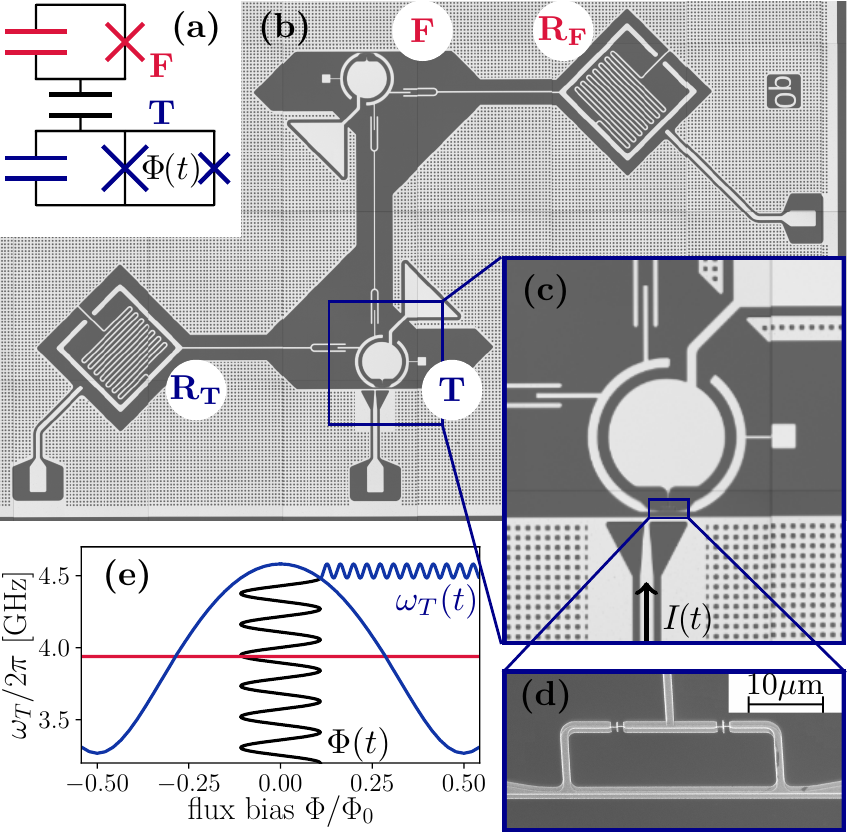}
\caption{Two-qubit circuit consisting of two capacitively coupled transmon qubits, one at fixed frequency~(F) and one tunable~(T).
(a) Lumped element circuit corresponding to our coupled qubits, where the tunable transmon is composed of an asymmetric SQUID.
(b) Optical image of the chip. Lumped element readout resonators are labeled $\mathrm{R}_\mathrm{F}$ and $\mathrm{R}_\mathrm{T}$.
(c) Time-varying flux pulses are actuated by a current $I(t)$ applied through the flux-bias line near T.
(d) Scanning electron micrograph of the SQUID loop of the tunable transmon.
(e) First transition frequency of the flux-tunable transmon (blue) and fixed-frequency transmon (red). Biasing the
flux near the maximum or minimum of the band ensures long coherence times by removing first order sensitivity to flux noise. The effect of flux modulation $\Phi(t)$ on the transmon frequency $\omega(t)$ is also shown in blue.
\label{fig:photo-circuit-bands}}
\end{figure}

Our experimental results were obtained with the circuit shown in Fig.~\ref{fig:photo-circuit-bands}(a). This circuit comprises a fixed-frequency transmon ($\mathrm{F}$) and a tunable transmon ($\mathrm{T}$), coupled capacitively with a strength $g$.
The tunable transmon, based on an asymmetric SQUID, has a first transition frequency $\omega_T$ that depends on the externally applied flux bias $\Phi$ as shown in Figure~\ref{fig:photo-circuit-bands}(e).
We modulate $\omega_T$ by applying a flux bias of the form
\begin{equation}\label{eq:flux-mod}
\Phi(t)=\phib + \phit \cos(\wpump t+\theta_p),
\end{equation}
with static flux bias $\phib$ and flux modulation amplitude $\phit$. An important special case occurs when the parking flux $\phib$ is set at a turning point in $\omega_T(\Phi)$, where the frequency modulation takes the approximate form
\begin{equation}\label{eq:freq-mod}
\omega_T(t) \approx \wb(\phit) + \wt(\phit) \cos(2\wpump t+2\theta_p)
\end{equation}
for $\phit \lesssim \Phi_0/2$.
The factors of 2 in Equation~\ref{eq:freq-mod} arise because, at these turning points, the transmon frequency undergoes two cycles for each cycle of the flux.
The nonlinearity of $\omega_T(\Phi)$ results in the time-averaged frequency $\wb$ being shifted away from $\omega_T(\phib)$ by an amount that depends on $\phit$.
Our experimental results were obtained with the tunable transmon parked at $\phib=0$, as depicted in Fig.~\ref{fig:photo-circuit-bands}(e). This condition offers first-order suppression of decoherence due to flux noise while idling~\cite{VionScience02}.

\begin{figure*}
\includegraphics[width=\textwidth]{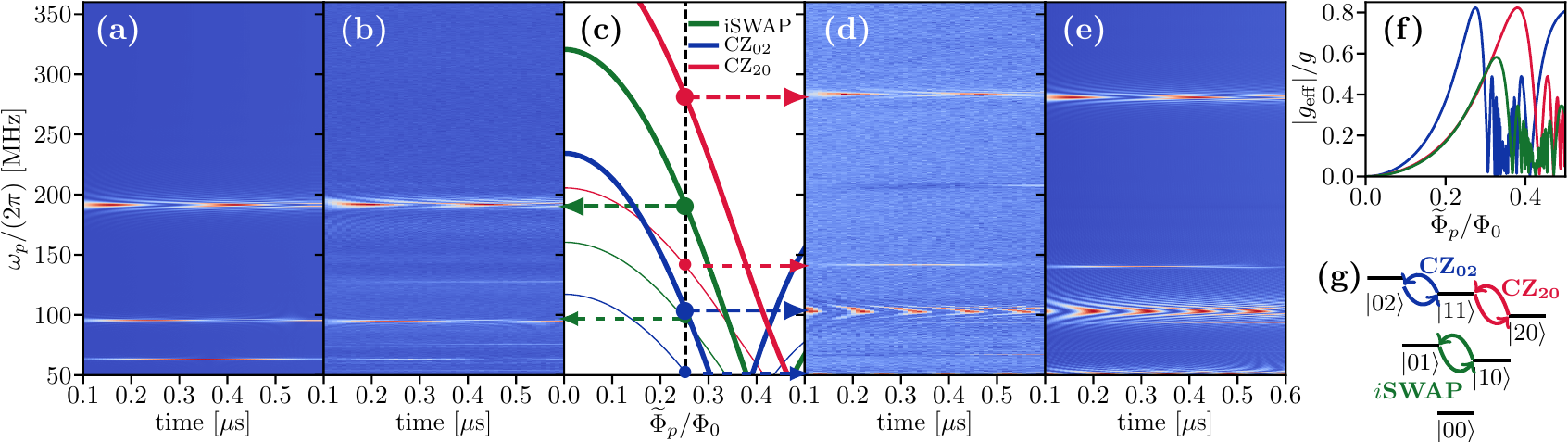}
\caption{Rabi oscillations driven by flux pulses. With a constant modulation amplitude and state preparation, we sweep pulse frequency and duration. The RF flux drive acts as a pump that compensates the energy detuning between two states. Rabi oscillations are observed when the modulation frequency matches the detuning between a pair of two-qubit states. With $\ket{10}$ prepared (a and b) we observe $\ket{10} \leftrightarrow \ket{01}$ swapping, and with $\ket{11}$ prepared (d and e) we observe both $\ket{11} \leftrightarrow \ket{02}$ and $\ket{11} \leftrightarrow \ket{20}$ swapping. The resonant frequencies depend on the modulation amplitude as shown in thick lines in (c), while the first harmonics of the resonances are shown in thin lines. Together these plots show a concordance between theory (c), dynamical simulations (a and e), and data (b and d) obtained with a modulation amplitude of $0.245 \Phi_0$. In (f) we show the effective coupling strength $g_{\mathrm{eff}}$ as a function of modulation amplitude for each transition, and in (g) we show the energy level diagram of the two-qubit system.
\label{fig:birch-plot}}
\end{figure*}

The effect of the parametric modulation can be seen in an interaction frame defined by the instantaneous qubit frequencies. Indexing the two-transmon excited states as $\ket{FT}$ and approximating the (defined positive) transmon anharmonicities $\eta_F$ and $\eta_T$ as constant, we write the interaction Hamiltonian
\begin{align}
\hat{H}_\mathrm{int}=\;& g\! \sum_{n=-\infty}^\infty \BJ_n\!\left(\frac{\wt}{2\wpump}\!\right) e^{i(2n\wpump t+\beta_n)}\nonumber\\
\times\;& \Big\{ \Exp{-i\Delta t}\ket{10}\bra{01}\nonumber\\
+\;& \sqrt{2} \Exp{-i(\Delta+\eta_F)t}\ket{20}\bra{11}\nonumber\\
+\;&\sqrt{2} \Exp{-i(\Delta-\eta_T)t}\ket{11}\bra{02}+\mathrm{h.c.}\Big\},
\label{Hint}
\end{align}
where $\BJ_n$ is the $n^{\mathrm{th}}$ Bessel function of the first kind and where we denote the detuning and phase
\begin{align}
\Delta =\, &\wb(\phit)-\omega_F \label{eq:shifted-detuning}\\
\beta_n =\, &(\wt/2\wpump)\sin(2\theta_p)+(2\theta_p+\pi)n \label{eq:coupling-phase}.
\end{align}
The interaction Hamiltonian makes plain the resonance conditions
\begin{align}
2n\wpump =\, & \Delta(\phit) \quad &\ket{10} \leftrightarrow \ket{01} \quad &i\mathrm{SWAP} \label{eq:iSWAPresonance} \\
2n\wpump =\, & \Delta(\phit)-\eta_T \quad &\ket{11} \leftrightarrow \ket{02} \quad &\mathrm{CZ}_{02} \label{eq:CZ02resonance} \\
2n\wpump =\, & \Delta(\phit)+\eta_F \quad &\ket{11} \leftrightarrow \ket{20} \quad &\mathrm{CZ}_{20} \label{eq:CZ20resonance}
\end{align}
with harmonics $n = \pm1, \pm2, \pm3, \dots$. These three types of entangling interactions are thus available at a series of amplitude-dependent modulation frequencies corresponding to the level spacings in the driven two-qubit system.
Each interaction has an effective coupling strength $g_\mathrm{eff}^{(n)}$ that determines the Rabi frequency and resonant linewidth of the interaction at the $n^{\mathrm{th}}$ harmonic.
This is given by the time-independent prefactor of each term in the Hamiltonian, so that
\begin{align}
g_\mathrm{eff}^{(n)} & = g\BJ_n\!\left(\frac{\wt}{2\wpump}\!\right)  \quad i\mathrm{SWAP} \\
g_\mathrm{eff}^{(n)} & = \sqrt{2} g\BJ_n\!\left(\frac{\wt}{2\wpump}\!\right)  \quad \mathrm{CZ}.
\end{align}
The flux drive introduces phase shifts $\beta_n$ to the Hamiltonian, which depend on both the amplitude and phase of modulation. The amplitude-dependent term in Equation~\ref{eq:coupling-phase} can be made negligible by the use of pulses with rising and falling edges much slower than $2 \omega_p$. The Rabi frequency is maximal, and the gate time minimal, where $\BJ_n$ is maximal. For the first harmonic this occurs for $\wt/2\wpump \approx 1.84$, giving $\BJ_1 \approx 0.582$.

The interaction Hamiltonian was derived under a rotating wave approximation, neglecting couplings between highly detuned levels. This approximation is valid for modulation frequencies~$2\omega_p$ and couplings~$g$ well below the qubit frequencies. We also neglected the modulation of the anharmonicity and have kept only the leading term in the coupling strengths between the different transitions. Taking these effects into account leads to small adjustments in the effective coupling rate for the range of modulation amplitudes used in this work. A more comprehensive analysis of this interaction is presented in Ref.~\cite{TheoryPaper}.

A parametric drive that resonantly couples two levels produces swapping in the subspace of those levels, described by the unitary
\begin{equation}
\label{eq:unitary}
\hat{U} = \begin{pmatrix}
\cos(\theta/2) & ie^{-i\phi}\sin(\theta/2) \\
ie^{i\phi}\sin(\theta/2) & \cos(\theta/2)\,
\end{pmatrix},
\end{equation}
with the population exchange given by $\theta = 2\int_0^\tau g_{\mathrm{eff}}(t)\, \mathrm{d}t$ during the flux pulse, and the phase $\phi$ of the exchange given by $\beta_n$. The $i$SWAP gate, mapping $\alpha\ket{00}+\beta\ket{01}+\gamma\ket{10}+\delta\ket{11}$ to
$\alpha\ket{00}+i\gamma\ket{01}+i\beta\ket{10}+\delta\ket{11}$, is realized by selecting a modulation amplitude and frequency to satisfy the $i$SWAP condition~(Equation \ref{eq:iSWAPresonance}) and pulsing the modulation for a time $\tau$ that yields $\theta=\pi$.
The CZ gates, mapping $\alpha\ket{00}+\beta\ket{01}+\gamma\ket{10}+\delta\ket{11}$
to $\alpha\ket{00}+\beta\ket{01}+\gamma\ket{10}-\delta\ket{11}$, are realized by activating one of the CZ resonances (Equations~\ref{eq:CZ02resonance} and~\ref{eq:CZ20resonance}) for a time that yields $\theta=2 \pi$. Physically, this causes the $\ket{11}$ state to fully leave and return to the computational basis, picking up the $SU(2)$ phase of -1 in the process. Corrections of local phases, necessary to realize the correct unitary for each gate, are discussed below.

\section{Experimental results}
We implemented this proposal on a two-transmon device fabricated on a high-resistivity ($>$10 k$\Omega$-cm) silicon substrate. In successive steps, Ti/Pd alignment marks and large Al features (ground planes, signal lines, and compact LC readout resonators) were defined using photolithography, electron-beam deposition, and liftoff. The transmon qubits, including Josephson junctions, were fabricated using the controlled undercut technique~\cite{Lecoq}. A bilayer resist stack consisting
of MMA (8.5) MAA copolymer and PMMA 950 was written with a 100-kV electron-beam lithography tool, and Al films were deposited in a Plassys MEB 550S at opposite substrate angles with an intervening controlled oxidation to form tunnel junctions.

\begin{figure*}
\includegraphics[width=0.8\textwidth]{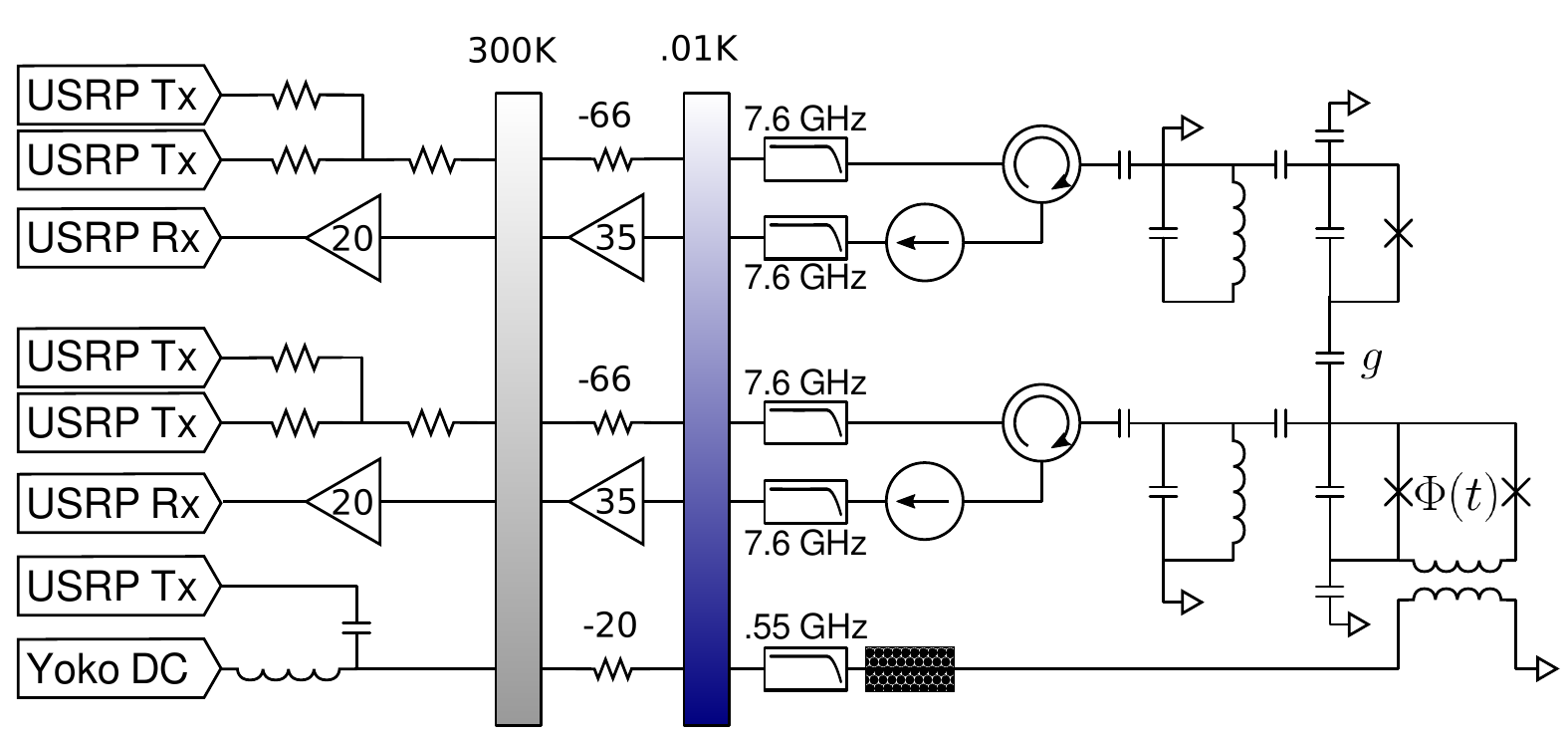}
\caption{Schematic of signal delivery to the two-qubit chip studied in this paper. Qubit drive and readout tones were transmitted, and the readout tone received, by Ettus Research USRP modules running with X300 daughterboards. The combined slow- and fast-flux components were generated by a Yokogawa GS200 DC power supply and another USRP X300, attenuated by 20dB at 4K and delivered through a UHF low-pass filter and a custom, dissipative Eccosorb filter. The on-chip flux-bias line terminated in an inductive short to ground.
\label{fig:expt-diagram}}
\end{figure*}

The transmons were capacitively coupled to each other with a strength of $g/2\pi=6.3\units{MHz}$, and to separate linear readout resonators with strengths of $\approx70\units{MHz}$. The microwave $XY$ control signals were delivered through the readout resonators, and the $Z$ control of the tunable transmon was delivered on a single-ended, on-chip flux-bias line.
All RF signals used for qubit control and readout were transmitted (Tx) and received (Rx) by the Ettus Research USRP X300 software-defined radio platform, modified with custom gateware.
A schematic of the experiment is given in Fig.~\ref{fig:expt-diagram}.
The tunable transmon frequencies were $\omega_{T}^\mathrm{max}/2\pi=4582\units{MHz}$, $\eta_T^\mathrm{max}/2\pi=173\units{MHz}$,
$\omega_{T}^\mathrm{min}/2\pi=3285\units{MHz}$, $\eta_T^\mathrm{min}/2\pi=185\units{MHz}$.
The fixed transmon frequencies were $\omega_F/2\pi=3940\units{MHz}$ and $\eta_F/2\pi=180~\mathrm{MHz}$.

The chip was operated with the tunable transmon parked at $\omega_T^\mathrm{max}$, where its relaxation and coherence times were $T_1=13\units{\mu s}$ and $T_2^*=10\units{\mu s}$. The fixed transmon had $T_1=34\units{\mu s}$ and $T_2^*=20\units{\mu s}$. Single-shot readout assignment fidelities of 0.85 (fixed) and 0.92 (tunable) were achieved with dispersive readouts combined with a binary classifier trained on $\ket{0}$ and $\ket{1}$ state preparation for each qubit~\cite{WhitePaper}. This classifier was used to classify the simultaneous readout signals obtained in the two-qubit tomography and randomized benchmarking results described below. Single-qubit randomized benchmarking decay constants of greater than 0.98 were obtained.
Application of the flux modulation increased the tunable transmon's sensitivity to flux noise, such that we measured $T_2^* \approx 4\units{\mu s}$ with the drive applied. In principle the flux modulation may also increase the qubit's relaxation rate~\cite{MartinisPRL05,Kelly2015}, but we found that $T_1$ was consistent with the flux drive on versus off.

As illustrated in Fig.~\ref{fig:birch-plot}, we observed the resonance conditions for the $i$SWAP and CZ interactions as predicted by theory. To obtain this concordance we measured the transfer function of flux pulses to the tunable transmon by measuring the average qubit frequency $\wb$ during modulation, as a function of both amplitude and frequency over the nominal $0$--$500\units{MHz}$ passband of the signal chain. For data-taking we then applied a frequency-dependent amplitude correction, similar to~\cite{Schuster}, to flatten the frequency response.

\begin{figure}[]
\includegraphics[width=\columnwidth]{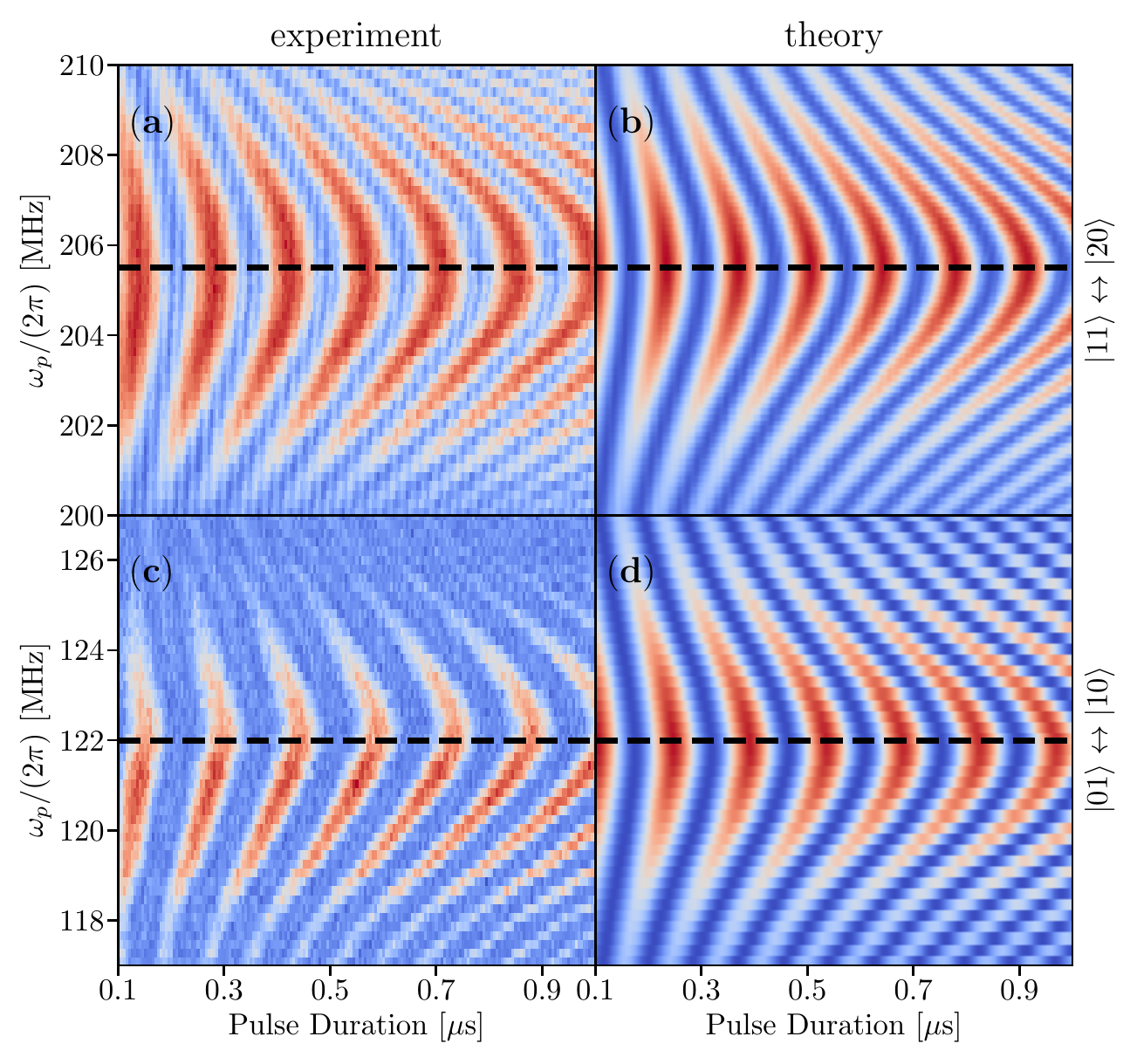}
\caption{A zoomed-in view of resonant features, similar to those shown in Fig.~\ref{fig:birch-plot}, reveals a ``chevron" pattern characteristic of Rabi oscillations. For each type of gate, a modulation amplitude was chosen to provide a suitable $g_{\mathrm{eff}}$. With that amplitude held constant, the flux-pulse duration and frequency were varied and values were chosen to give the roughly the desired gate up to local phases. The patterns observed in the experimental data (left) were in good agreement with the predictions of a dynamical simulation (right), as shown here for transitions $\ket{01} \leftrightarrow \ket{10}$ (bottom) and $\ket{11} \leftrightarrow \ket{02}$ (top). The only free parameters in the model are the effective relaxation and coherence times in the presence of flux drives.\label{fig:gate-chevrons-zoom}}
\end{figure}

Closer inspection of the resonant features also showed good agreement between theory and observations of a characteristic ``chevron" signal (see Fig.~\ref{fig:gate-chevrons-zoom}). This signal was used to confirm the resonant frequency at the chosen value of modulation amplitude. The parametric $i$SWAP and CZ gates were then realized by determining which pulse durations $\tau$ yielded the appropriate values of $\theta$.

Once a fully entangling interaction is specified, realizing the desired unitary for each gate requires a number of corrections to be applied.
\begin{enumerate}
    \item The mean qubit frequencies are shifted during the modulation and thus acquire local phases with respect to their microwave drives. These local phases were measured with a Ramsey experiment where only the phase of the second $\pi/2$ pulse was varied. The corrections for the phases were then implemented as local updates to the phases of downstream microwave pulses.
    \item The frequency of the flux pulse must be chosen according to the resonant condition (Equation~\ref{eq:shifted-detuning}), which is affected by the shifts in qubit frequencies. However, when the flux pulse is not applied, the dynamical phase between the two levels accumulates at the unshifted detuning of the two levels. Because the flux pulses were generated using a local oscillator tuned to the shifted frequency, the phases of successive pulses required correction for the mean shift of the qubit frequencies.
    \item The phase $\beta_n$ is present in the $i$SWAP interaction, where the off-diagonal elements of Equation~\ref{eq:unitary} are non-zero, and is set by the phase $\theta_p$ of flux modulation with respect to the individual qubit frames. This interaction phase can be handled either by updating the phase of the flux drive or pushed to equal and opposite local frame changes applied after the gate.
    \item The exchange of states in the $i$SWAP gate entails the exchange of their local phases. This was tracked by swapping the microwave phases after each $i$SWAP gate.
\end{enumerate}

\begin{table}[]
    \centering
    \begin{tabular}{c|c|c|c}
        \multicolumn{1}{l|}{Gate}                        & $i$SWAP   & CZ$_{02}$ & CZ$_{20}$  \\
        \hline\hline
        \multicolumn{1}{l|}{$\widetilde{\Phi}$ [$\Phi_0]$} & 0.317   & 0.245     & 0.280    \\
        \multicolumn{1}{l|}{$f_p$ [MHz]}                 & 122       & 112       & 253          \\
        \multicolumn{1}{l|}{duration [ns]}               & 150       & 210       & 290        \\
        \multicolumn{1}{l|}{$g_\mathrm{eff}/2\pi$ [MHz]} & 3.9       & 4.0       & 2.4        \\
        \multicolumn{1}{l|}{Effective $T_1$ [$\mu$s]}    & 5-25      & 7-17      & 7-17       \\
        \multicolumn{1}{l|}{Effective $T_2^*$ [$\mu$s]}  & 4.6(8)    & 10.2(22)  & 7.1(11)    \\
        \multicolumn{1}{l|}{IRB fidelity}                & 0.94      & 0.92      & 0.91      \\
        \multicolumn{1}{l|}{- Clifford fidelity}         & 0.94      & 0.93      & 0.88    \\
        \multicolumn{1}{l|}{QPT fidelity}                & 0.93      & 0.92      & 0.92       \\
        \multicolumn{1}{l|}{- unitarity bound}           & 0.939     & 0.945     & 0.935    \\
        \multicolumn{1}{l|}{- interferometric bound}     & 0.938     & 0.945     & 0.935    \\
    \end{tabular}
    \caption{Summary of experimental results. The operating point for each gate is given by the flux-modulation amplitude and frequency. The gate durations and $g_\mathrm{eff}$ values include the pulse risetimes of 30-40 ns (to suppress the effect of pulse turn on phase). The values of $T_1$ and $T_2^*$ were estimated under drive conditions by measuring the decay of the qubit as a function of flux-pulse duration. Fidelities are given using both process tomography and randomized benchmarking, with accompanying bounds on the tomography result. Further detail on the fidelity characterization is provided in the main text.}
    \label{tab:gate_points}
\end{table}

\begin{figure}
\includegraphics[width=\columnwidth]{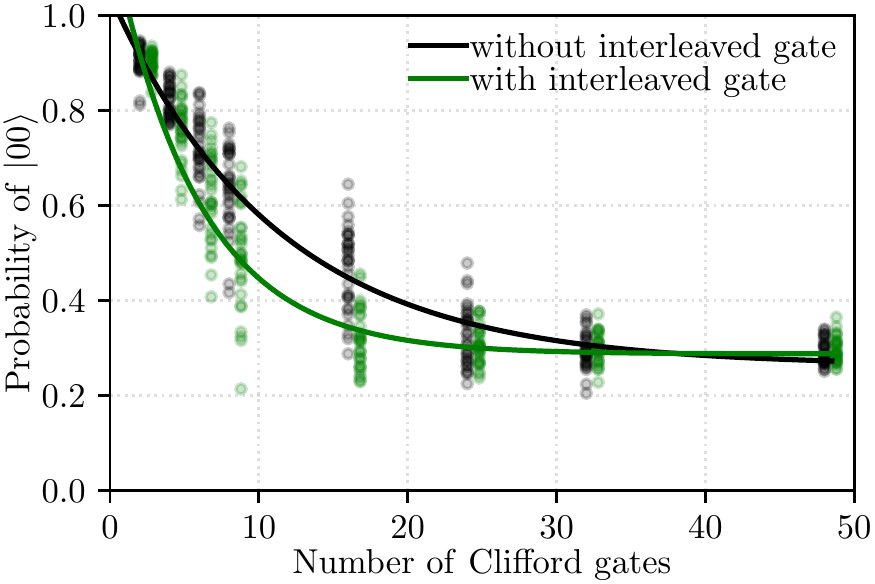}
\caption{Interleaved randomized benchmarking result for the $i$SWAP gate. In this experiment, sequence lengths of $\{2, 4, 6, 8, 16, 24, 32, 48\}$ were used. For visual clarity, the data points for each sequence length are offset symmetrically about the true length. The fit curves, however, are not offset. The accelerated decay from the expected final state of $\ket{00}$ is attributed to the error in the $i$SWAP.
\label{fig:irb}}
\end{figure}

\begin{figure*}
\includegraphics[width=0.9\textwidth]{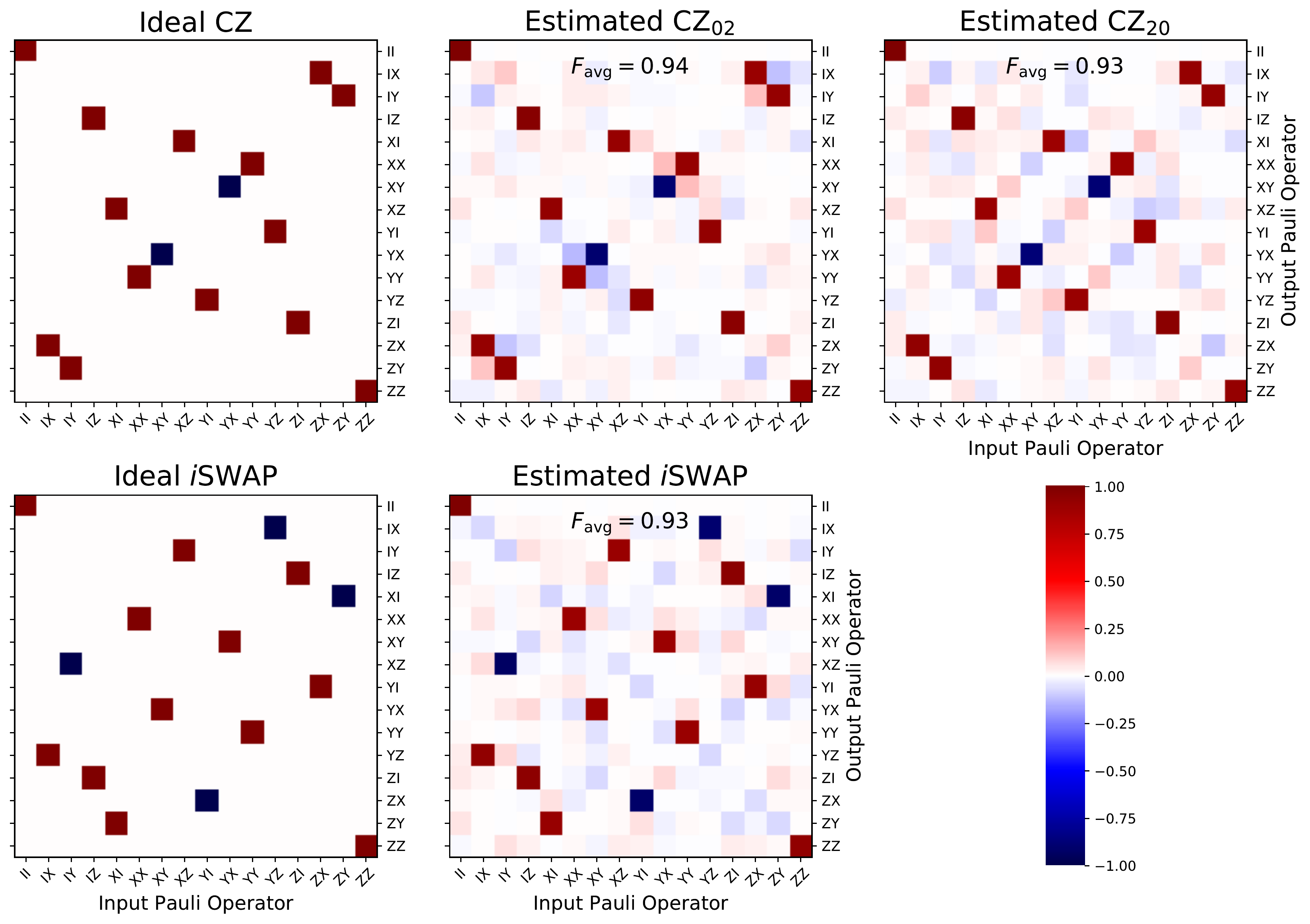}
\caption{Two-qubit process tomography parametric CZ (top) and $i$SWAP (bottom) processes. The corresponding average process fidelities are $F=0.92$ for both CZ's, and $F=0.93$ for the $i$SWAP.
\label{fig:tomography}}
\end{figure*}

Characteristics of each gate are given in Table~\ref{tab:gate_points}.
The average gate fidelity of each operation was estimated using interleaved randomized benchmarking (IRB)~\cite{magesan2012-interleaved-rb} as well as maximum-likelihood quantum process tomography (QPT)~\cite{hradil}.
In using QPT to estimate the gate fidelities~\cite{Horodecki1999,Nielsen2002} we compensated for the known readout infidelities~\cite{Ryan2015,WhitePaper} and applied complete-positivity (CP) constraints.
We estimate $F_{i\text{SWAP}}=0.93$ and $F_{\text{CZ}}=0.92$ for both CZ$_{\mathrm{02}}$ and CZ$_{\mathrm{20}}$.
The constrained reconstructions, along with the ideal target gates, are depicted in Fig.~\ref{fig:tomography}.
Similar devices have also yielded average fidelities for CZ above $0.9$~\cite{WhitePaper}. The result of the IRB experiment for the $i$SWAP gate is shown in Fig.~\ref{fig:irb}.

In order to roughly identify the source of the observed infidelities, we estimate the fidelity to
the unitary closest to the reconstructed evolution. If the fidelity to the closest unitary is similar to the fidelity
to the ideal target gate, we conclude there were no significant coherent errors, and that the fidelity is limited
by the decoherence of the evolution. Otherwise, we conclude there are significant coherent errors in the evolutions.
There are two natural notions of closest unitary for a reconstructed superoperator $\mathcal E$ with corresponding
canonical Kraus representation $\mathcal E(\rho)=\sum_i K_i \rho K_i^\dagger$~\cite{Havel} and with Liouville representation
singular value decomposition $\mathcal E = \mathcal U\mathcal S\mathcal V^\dagger$. One is given by
the unitary $U_0$ where $\|K_0\|\ge\|K_{i>0}\|$, and where $K_0=U_0 P_0$ is a polar decomposition~\cite{Oi}, which results in the approximate
interferometric bound
\begin{equation}
\overline F \lessapprox \frac{\mathrm{tr}~(U_0^*\otimes U_0)^\dagger \mathcal{E} + d}{d^2+d}.
\end{equation}
The other is given by $\mathcal U\mathcal V^\dagger$, which results in the strict Procrustean unitarity bound
\begin{equation}
\overline F \le \frac{\mathrm{tr}~\mathcal{S} + d}{d^2+d}.
\end{equation}
These upper bounds were consistent with our measured fidelity estimates, indicating that infidelities are largely explained by the decoherence during the gate evolution.

\paragraph{Summary ---}
We have proposed and demonstrated parametric $i$SWAP and CZ gates activated by modulating the frequency of a tunable transmon. We have shown close agreement between observed and predicted resonance conditions over a wide range of frequencies and multiple drive amplitudes, and we have described and implemented local phase corrections necessary to produce correct and repeatable unitaries. Gate fidelities, estimated by interleaved randomized benchmarking and quantum process tomography, were estimated to be 0.91-0.94 and appeared to by limited by decoherence during the parametric drive. The underlying interactions used in this technique are highly selective and provide a way to mitigate frequency crowding in a scalable quantum processor architecture.

\begin{acknowledgments}
Part of this work was performed at the Stanford Nano Shared Facilities (SNSF), supported by the National Science Foundation under award ECCS-1542152.
\end{acknowledgments}

\bibliography{pgates.bib}

\end{document}